\documentstyle[12pt]{article}

\newcommand{\nc}{\newcommand}
\nc{\beq}{\begin{equation}}
\nc{\eeq}{\end{equation}}
\nc{\beqa}{\begin{eqnarray}}
\nc{\eeqa}{\end{eqnarray}}
\newcommand{\mysection}[1]{\setcounter{equation}{0}\section{#1}}

\input epsf
\newwrite\ffile\global\newcount\figno \global\figno=1

\def\writedef#1{}
\def\figin{\epsfcheck\figin}\def\figins{\epsfcheck\figins}
\def\epsfcheck{\ifx\epsfbox\UnDeFiNeD
\message{(NO epsf.tex, FIGURES WILL BE IGNORED)}
\gdef\figin##1{\vskip2in}\gdef\figins##1{\hskip.5in}% blank space instead
\else\message{(FIGURES WILL BE INCLUDED)}%
\gdef\figin##1{##1}\gdef\figins##1{##1}\fi}
\def\figinsert{}
\def\ifig#1#2#3{\xdef#1{fig.~\the\figno}
\writedef{#1\leftbracket fig.\noexpand~\the\figno}%
\figinsert\figin{\centerline{#3}}\medskip\centerline{\vbox{\baselineskip12pt
\advance\hsize by -1truein\center\footnotesize{  Fig.~\the\figno.} #2}}
\bigskip\endinsert\global\advance\figno by1}
\def\endinsert{}
\begin{document}

\title{\large{\bf  Lattice Tests of Supersymmetric Yang-Mills 
Theory?}}

\author{  
%Yale Bulldogs and BU Terriers
%\\
Nick Evans\thanks{nevans@physics.bu.edu}
\\ \\ Department of Physics, Boston University, Boston, MA 02215 \\ \\
Stephen D.H.~Hsu\thanks{hsu@hsunext.physics.yale.edu},
Myckola Schwetz\thanks{ms@genesis2.physics.yale.edu}
\\ \\ Department of Physics, Yale University, New Haven, CT 06520-8120 \\
}

\date{July, 1997}

\maketitle

\begin{picture}(0,0)(0,0)
%\put(350,370){draft version}
\put(350,350){BUHEP-97-20}
\put(350,335){YCTP-P11-97}
%\put(350,385){hep-th/97?????}
\end{picture}
\vspace{-24pt}

\begin{abstract}
Supersymmetric Yang Mills theory is directly accessible to 
lattice simulations using current methodology, and can 
provide a non-trivial
check of recent exact results in SQCD. In order to tune the lattice
simulation to the supersymmetric point it is neccessary to 
understand the behavior of the theory with a small supersymmetry 
breaking gaugino mass. We introduce a soft breaking gaugino
mass in a controlled fashion using a spurion analysis.
We compute the gluino condensate, vacuum energy 
and bound-state masses as a function of the gaugino mass, 
providing more readily accessible predictions which still 
test the supersymmetric results. 
Finally we discuss diagnostics for obtaining 
the bare lattice parameters that correspond to the 
supersymmetric continuum limit.

\end{abstract}

\newpage
\section{Introduction}

The recently proposed exact solutions \cite{seiberg1,seiberg2,seiberg3}
of N=1 supersymmetric QCD (SQCD) satisfy
striking tests of self-consistency and provide an extremely plausible
picture of the rich low-energy dynamics of these models.  
Nevertheless, one may feel a little discomfort at the absence of direct 
non-perturbative tests of the results.
An obvious possibility is that these models could be simulated directly 
on the lattice. Some initial work in these directions has already been
performed in \cite{Montvay}.
Unfortunately, as is well known, lattice regularization violates 
supersymmetry \cite{CV}, and a special fine-tuning is required to
recover the SUSY limit (this is analogous to the case of chiral 
symmetry in lattice QCD). Away from the SUSY point, the continuum
limit of the lattice theory is described by a model with explicit SUSY  
violating
interactions. In some cases, these violations may correspond only
to soft breakings \cite{giradello}, 
although this is not guaranteed in general. 

Softly broken SUSY models can be studied using spurion techniques,
and ``exact'' results are possible \cite{soft1,soft2} (additional
investigations of softly broken SQCD can be found in \cite{soft3}).
In this paper, we propose some tests of supersymmetric Yang-Mills
theory (SYM -- or SQCD with zero flavors of quark) 
which can be carried out using lattice techniques.
SYM is a simple theory with only one parameter,
the gauge coupling. The only low-dimension (renormalizable) 
SUSY violation allowed by gauge invariance is a gaugino mass,
which is a soft violation. Therefore, the continuum limit of the
lattice regularized version of SYM is simply SYM with a massive gaugino.
The SUSY limit can be reached by fine-tuning the lattice parameter
corresponding to a bare gaugino mass. 
In order to understand this limit as well as possible, we study continuum
SYM with explicit gaugino mass, and derive some relations describing
the approach to the SUSY limit. The spurion techniques used assume
that there is not a phase transition to some totally new phase 
the moment that supersymmetry is broken, a fact that may be tested on
the lattice (if such a transition did occur then testing the
supersymmetric predictions would become very hard since that phase would 
constitute only a single point in parameter space and there would be
no understanding of the approach to that point). 
Assuming no such transition, several non-trivial predictions can
be made regarding the vacuum energy and of the behavior of the 
gaugino condensate. A less rigorously derived description of the 
lightest bound states of SYM theory has also been proposed in the 
literature \cite{VY,softVY} from which predictions for the masses 
of the gluino-gluino and glue-gluino bound states and their splittings
away from the supersymmetric point may be obtained.
We also discuss aspects of tuning toward the SUSY limit.

A lattice test of the pure glue theory would also provide 
a test of the tower of SQCD theories with fundamental matter flavors. 
The numerical coefficients in the
effective superpotential of SQCD theory with $N_f =
N_c-1$ have been determined analytically \cite{Cordes} 
(this is the first value of $N_f$ where the 
gauge group can be broken completely
and the theory studied in the perturbative regime). The pure glue theory
predictions resulting from integrating out the quark flavors in that
effective theory  are closely related to these coefficients 
as we review below. 

The outline of the paper is as follows. In section 2 we review the
holomorphic analysis of SYM. In sections 3 and 4 we introduce 
soft breakings into the analysis, deriving predictions for the lattice
simulations. In section 5 we discuss diagnostics for tuning towards 
the SUSY point. In section 6 we summarize our results. Finally, in the
appendix, we discuss the rescaling anomaly and the holomorphic
vs. canonical field normalizations, which are relevant to the comparison
between exact and lattice results.

\section{SUSY Yang-Mills}

The bare Lagrangian of  SYM  with $SU(N_c)$ gauge group  
is
\begin{equation}
\label{SYM}
{\cal L} ~=~ \frac{1}{g_0^2}\left[ \,-\frac{1}{4} 
G_{\mu\nu}^a G_{\mu\nu}^a
~+~ i\lambda_{\dot\alpha}^\dagger D^{\dot\alpha\beta}\lambda_\beta
\right] ~. 
\end{equation}
This model possesses a discrete global $Z_{2N_c}$ symmetry, a 
residual non-anomalous subgroup of the anomalous chiral $U(1)$. The theory 
is believed to generate a gaugino condensate breaking the $Z_{2N_c}$ 
symmetry to $Z_{2}$ and also to exhibit a mass gap due to confinement.

In supersymmetric notation the Lagrangian (\ref{SYM}) can be written as
\beq
\label{SSYM}
{\cal L} ~=~ \int d^2 \theta~
\frac{1}{8\pi} Im\, \tau_0 W^{\alpha}W_{\alpha}~, 
\eeq
where the gauge coupling is defined to be
$\tau_0 = \frac{4\pi i}{g^2_0} + \frac{\theta_0}{2\pi}$.

In this notation, the strong coupling scale of SYM is defined through 
the relation
\beq
\label{SL}
\Lambda ~\equiv~e^{2\pi i \tau_0/ b_0} \Lambda_{UV}~, 
\eeq
where $b_0 = 3N_c$ is the first coefficient of SYM $\beta$-function and
$\Lambda_{UV}$ is the UV cut off of the theory at which the coupling takes
the value $g_0$. The relation of this definition of the strong scale to 
the $\rm \bar{MS}$  scheme has been investigated in \cite{Pouliot}.
Note that $\Lambda^{b_0}$ is explicitly $2\pi$-periodic in the
$\theta_0$-angle.
For the purposes of this paper we set $\theta_0 =0$.

To derive the low energy effective theory of SQCD we note that there are
two anomalous symmetries of the theory, $U(1)_R$ and scale
invariance. In
fact their anomalies are related since their currents belong to the same
supermultiplet. These symmetries can be restored in an enlarged theory
provided we allow the spurion couplings to transform:
\beqa
U(1)_R: \hspace {1cm} &&
W(x,\theta) \rightarrow e^{i\zeta} W(x, \theta e^{i\zeta}) \nonumber \\
&& \Lambda \rightarrow \Lambda e^{i2 \zeta/3} \nonumber 
\eeqa
\beqa
{\rm Scale} \hspace{0.1cm} {\rm  Invariance}: \hspace{1cm}
&& W(x,\theta) \rightarrow e^{3\xi/2} W(xe^\xi, \theta e^{\xi/2})
\nonumber \\
&& \Lambda \rightarrow \Lambda e^\xi ~~~~~.\nonumber
\eeqa

We may now determine the general form of the partition function 
(assuming a mass gap) in the large volume limit as a
function of $\tau$ subject to these symmetries. The only possible terms
are
\begin{equation}
\label{SZ}
Z[\tau] = {\rm} \hspace{0.1cm} {\rm exp} ~ i V \left[  {9 \over \alpha}  
\Lambda^\dagger \Lambda|_D - ( \beta
\Lambda^3|_F +h.c.) \right]
\end{equation}

The numerical coefficients $\alpha$ and $\beta$ remain undetermined from
the above symmetry arguments. $\beta$ may be determined from the results
for SQCD with massive quarks. The $U(1)_R$ symmetries of models with
matter are sufficient to determine the form of the superpotential as
\begin{equation}
W_{\rm massive} = c~ \left[ {\rm det}(m) \Lambda^{b_0} \right]^{1/ N_c}
\end{equation}
where $\Lambda$ is defined similarly to (\ref{SL}) but with the
appropriate $\beta$-function $b_0 = 3N_c-N_f$ and $c$ is a constant,
undetermined by the symmetry. The scale transformation anomaly is not
sufficient to uniquely determine the K\"{a}hler potential in these theories
with two mass scales $\Lambda$ and $m$. The coefficient $c$ may
be explicitly calculated in theories in which $N_f \geq N_c-1$ since
the squark vevs may be chosen to completely break the gauge symmetry in
the perturbative regime. The superpotential term is then generated by
calculable instanton effects. The calculation was performed in
\cite{Cordes}. The mass terms may then be taken to infinity removing the
quarks from the theory and leaving the pure glue theory in a controlled
fashion. The resulting prediction for $c$ in Minkowski space 
is $(N_f-N_c)$ and hence
$\beta$ is $N_c$.

These strong arguments lead to two predictions for the
condensates of the SYM theory. The source $J$ for the
gaugino correlator $\lambda \lambda$ occurs in the same position as the 
F-component of $\tau$ and is hence known. 
There are two independent correlators
\begin{eqnarray}
\label{cond}
\langle \lambda \lambda \rangle  & = &  -32 \pi^2 \Lambda^3\nonumber \\
\langle \bar{\lambda} \bar{\lambda} \lambda \lambda\rangle &  = & 
{- 1024 i \pi^4 \over \alpha N_c^2} |\Lambda|^2 / V~~~.
\end{eqnarray}

The IR theory has a gaugino condensate $\simeq \Lambda^3$, with phase
$2\pi i\tau /N_c$ and hence there are $N_c$ degenerate 
vacua associated
with the $N_c$th roots of unity. Below, therefore, $\Lambda^3$ 
is an $N_c$ valued constant with phases $n 2\pi i /N_c$ where $n$ runs from
$0...$ $N_c-1$.

\section{Soft Supersymmetry Breaking}

We may induce a  bare gaugino mass through a non zero
F-component of the bare gauge coupling 
$\tau = \tau_0 + F_{\tau} \theta \theta$ 
\cite{soft1}
\begin{equation}
- {1 \over 8 \pi} Im [ F_{\tau} \lambda \lambda]
\end{equation}
To make the gaugino mass canonical we take $F_{\tau} = i 8 \pi m_\lambda$.
In the IR theory $\tau$ enters through the spurion, $\Lambda_s$,
the lowest component of which is the strong interaction scale
$\Lambda$  

\beq
\label{Lambda}
\Lambda^{b_0}_s=~\Lambda^{b_0}\, (1 ~-~ 16\pi^2 m_\lambda   \theta \theta )
\eeq 

$\Lambda$ occurs linearly in the superpotential of the
theory. Thus there will be a correction to the potential
of the form:
\beq
\label{correction}
\Delta V ~=~-32 \pi^2 Re ( m_\lambda  \Lambda^3)  - {256 \pi^4 
  \over \alpha N_c^2} |m_\lambda \Lambda|^2
\eeq
Terms with superderivatives acting on the spurion field
can also give rise to contributions to the potential but these are
higher order in an expansion in $m_\lambda/\Lambda$. The shift in
the potential energy of the $N_c$ degenerate vacua of the SYM theory
at linear order in $m_\lambda$ is known and we may determine the vacuum
structure 
\begin{equation}
\label{deltaV}
\Delta V = -32 \pi^2 |m_\lambda \Lambda^3|~ \cos \left[ {2 \pi n \over N_c} +
\theta_{m_\lambda} \right]
\end{equation}
For small soft 
breakings, $m_\lambda \ll \Lambda$, where the linear term dominates, 
the degeneracy between the SYM vacua is broken favoring one vacuum 
dependent on
the phase of the gaugino mass.
The coefficient in the energy shift
is a test of the exact superpotential in Eq.(\ref{SZ}).

We may also determine the leading shift in the gaugino condensate
\beq
\label{Stau}
\langle \lambda \lambda \rangle 
~=~  -32 \pi^2 \Lambda^3 \, 
~+~ \frac{512 \pi^4}{\alpha N^2_c} m_\lambda^* |\Lambda|^2~,
\eeq
which depends on the unknown parameter $\alpha$. Strictly speaking
there are also divergent contributions to this quantity which are
proportional to $m_\lambda$ times the cut-off squared. We will have
more to say about these divergences in section 5.

Reinserting the bare $\theta_0$ angle into the expression for the shift in
vacuum energy we find 
\begin{equation}
\Delta V = - 32 \pi^2 |m_\lambda \Lambda^3|~ \cos \left[ {2 \pi n \over N_c} +
\theta_{m_\lambda} + {\theta_0 \over N_c} \right]
\end{equation}
As $\theta_0$ is changed first order phase transitions occur at
$\theta_0 = ({\rm odd}) \pi$ where two of the $N_c$ SYM vacua
interchange as the minimum of the softly broken theory.

\section{The Lightest Bound States}

An alternative description of the low energy behaviour of SYM theory has
been presented by Veneziano and Yankelowicz \cite{VY} which attempts to
describe the lightest bound states of the theory. The form of their
effective action can be rigourously obtained from the discussion above.
Since the source $J$ for $WW$ occurs in the same places as the
coupling $\tau$ we also know the source dependence of $Z$. If we wish we
may Legendre transform $Z[\tau,J]$ to obtain the effective potential for
the classical field
\begin{equation}
S \equiv 
\frac{1}{32\pi^2}\,\mbox{Tr}\,\langle W^2 \rangle~~.
\end{equation}
We find 
\begin{equation}
\label{VY}
\Gamma[\tau,S] = {9 \over \alpha} \left( \bar S S \right)^{1/3}\Big|_D ~-~ 
N_c \left( S - S\ln (S/ \Lambda^3) \right)\Big|_F+ \, \mbox{h.c.}~.
\end{equation}
So derived, this effective action contains no more information than
Eq.(\ref{SZ}) simply being a classical potential whose minimum determines
the vev of $S$ and we find, by construction, Eq(\ref{cond}).

A stronger interpretation can also be given to the VY action. 
We can attempt to use it to describe the physical gluino-gluino
and gluino-glue bound states, if
we assert that they interpolate in the perturbative regime to the field $WW$.
Under this assumption the symmetry properties of the bound states 
would be those of the S field, reproducing the
VY action for those states, up to additional scale-invariant 
Ka\"hler terms. (The latter only affect our predictions at higher order in
the soft breaking.)
To obtain the physical states
one performs 
an appropriate rescaling of the $S$-field 
\beq
\label{Sres}
\Phi~=~ \frac{3}{\sqrt{\alpha}} \,S^{1/3}
\eeq
in the Lagrangian (\ref{VY}) 
to make the kinetic term canonical 

\begin{equation}
\label{VYLR}
{\cal L} ~=~ \left( \bar \Phi \Phi \right)\Big|_D~-~
\frac{\alpha^{3/2}N_c}{9}\left(\frac{1}{3}\Phi^3 -  \Phi^3 
\ln{(\frac{\alpha^{1/2}}{3}\frac{\Phi}{\Lambda} )}\right)\Big|_F
~+~ \, 
\mbox{h.c.}~.
\end{equation}

In fact, as discussed in \cite{Shifman}, this effective
Lagrangian is not complete since it does not possess the full $\rm Z_{2N_c}$
symmetry of the quantum theory. To restore that symmetry the extra term
\beq
\Delta {\cal L} = { 2 \pi i m \over 3} \left( S - \bar{S} \right)
\eeq
where $m$ is an integer valued Lagrange multiplier must be added. For
the $n=0$ vacuum with vanishing phase this extra term vanishes
and the VY model above is recovered. We shall concentrate on that vacuum and
real, positive mass perturbations which make that vacuum the true vacuum of 
the perturbed theory.  

In addition to the particle states associated with the $\phi$ field
(gluino--glunio or gluino--glue balls), one might expect to find
states associated with the familar glueballs of QCD, with interpolating
fields $F^2$ or $F \tilde{F}$, and their superpartners. 
A chiral supermultiplet can be constructed
out of $D^2 W^2 = D^2 S$ which contains the appropriate fields. We have
no reason to expect that the glueball states are parametrically heavier
than the $\phi$ field states, and so there is no {\it a priori} reason
to integrate them out of the effective Lagrangian for $\phi$. Furthermore,
any glueballs are likely to be strongly coupled to $\phi$ and have a
non-negligible effect on its dynamics.  

On the other hand, any couplings to glueballs are still constrained by
scale invariance and anomalous $U(1)_R$-invariance. Let us postulate
the existence of some effective glueball chiral 
superfield $\phi' (x)$, which is dimension
one and has a canonical kinetic energy term. Since $\phi' (x)$ is 
constructed from $D^2 W^2$ it must have R-charge zero. A superpotential
term describing the interaction of $\phi'$ with S would be of the form:
\beq
\int d^2 \theta ~ \Lambda^{3 - 3a - b} S^a \phi'^b ~~.
\eeq
The constraint from anomalous R-invariance (under which $\Lambda$ has
charge 2/3), requires that 
\beq
2 ~=~ 2a - 2/3 (3a + b - 3)~~,
\eeq
or $b = 0$. Therefore there are no superpotential terms of the correct
type. All interactions between the potential glueball excitations and
$S$ (including mass or wavefunction mixing) 
must appear as K\"{a}hler terms in the effective Lagrangian. There are
already unknown K\"{a}hler terms that enter into the physical masses. These
are higher derivative terms which, while suppressed by powers of
$\Lambda$, contribute to the wave function renormalization, $Z(p^2)$,
which must be evaluated at $p^2=m^2 \simeq \Lambda^2$ to obtain the
physical masses. The K\"{a}hler corrections from the $\phi'$ may be subsumed
into these unknown terms. 

Of course, it is possible that the glueball fields do not enter the effective
Lagrangian as $\phi' (x)$, but rather as some other effective field, or even
the auxilliary field of $S$ (which contains $G_{\mu \nu}^2$ and 
$G \tilde{G}$). In this case the description advocated here may not be 
correct. We will discuss this possibility (more specifically, the status of
glueballs in SYM) in more detail below, but for now we simply adopt the
VY Lagrangian and discuss its mass predictions.
A lattice simulation will hopefully test these predictions and shed light
on whether the action is indeed the correct description.

The straightforward evaluation of bosonic ($\lambda \lambda $) 
and fermionic ($g\lambda$) excitation masses  
around
the minimum from Eq(\ref{VYLR}) gives
\beq
\label{susymass}
m_{\lambda \lambda}~=~m_{g\lambda}~=~N_c \alpha \Lambda~.
\eeq

It is important to stress again that these masses are not the
physical masses of the bound states. Rather, they are zero-momentum
quantities, which are related to the physical ones by wave function
factors $Z(p^2 = m_{phys}^2)$. These wave function factors result from
higher-derivative K\"{a}hler terms in ${\cal L}$, and are unknown.

A soft breaking gaugino mass may again be introduced through the
F-component of the spurion $\Lambda_s$ (first investigated in \cite{softVY}). 
The new scalar potential is 
\beq
\label{newpot}
{\cal V}~=~ \frac{\alpha^3 N^2_c}{9} \,|\phi^2|^2\,
 \Big|\ln{(\frac{\alpha^{1/2}}{3}\frac{\phi}{\Lambda} )}\Big|^2 ~-~
~\frac{32\pi^2 \alpha^{3/2}}{27} Re\,m_\lambda \phi^3
\eeq
from which we can calculate the shifts in the masses of the bound
states. The two scalar fields and the fermionic field are all split in
mass. The eigenvalues of the mass matrices are
\begin{eqnarray}\label{mass}
M_{\rm fermion} & = & N_c \alpha \Lambda + {16 \pi^2 |m_\lambda| 
\over  N_c}  \nonumber \\
M_{\rm scalar} & = \nonumber & N_c \alpha \Lambda + {56 \pi^2 |m_\lambda| 
\over 3 N_c}\\
M_{\rm p-scalar} & = & N_c \alpha \Lambda + {40 \pi^2 |m_\lambda|  
\over 3 N_c}
\end{eqnarray}
these results have been derived for real, positive mass which favor the
supersymmetric vacuum with vanishing phase. For even $N_c$ there
is a supersymmetric vacuum characterized by $\langle S \rangle = - \Lambda^3$
which is prefered by negative, real mass perturbations. It is easy
to construct an effective lagrangian about that vacuum in a similar fashion to
(\ref{VYLR}). The bound state masses are again given by (\ref{mass}).

The physical masses are again related to these quantities by
unknown wave function renormalizations $Z$ which arise from  K\"{a}hler terms, 
$$
M_{\rm physical} 
~ \equiv ~ Z ~ M ~~. 
$$
Fortunately, we know that
in the SUSY limit the wavefunction factors are common
within a given multiplet. This degeneracy holds even after the vev of
the field is shifted by the soft breakings since a shift in the vev
alone (without SUSY breaking) leaves the physical masses degenerate
within a multiplet.
We also know that the {\it relative change} in these K\"{a}hler terms is of order 
${\cal O} (f_\tau^2)$, 
and hence can be ignored at leading order in
the soft breakings. Therefore, we may 
still obtain a prediction for the rate of change of
the ratios of the physical masses, 
\begin{eqnarray}
\label{bm}
\bar{M} (m_\lambda) & ~\equiv ~ & { Z (m_\lambda) M(m_\lambda )- Z(0) M(0)  
\over Z(0) M(0) }~~,
\nonumber\\
& \simeq & {\partial M \over \partial m_\lambda} 
\left[ \frac{1}{M}   + 
\frac{1}{Z}  {\partial Z \over \partial M } \right]  m_\lambda 
\end{eqnarray}
near the SUSY limit. 
The factor in brackets is common within a given multiplet.
Since the quantity $Z(m_0)$ is unknown, we can only predict
the {\it ratios} of $\bar{M}$ at the SUSY point or equivalently the
ratios
of $\partial M / \partial m_\lambda$
\begin{equation}
\partial M_{\rm ps} / \partial m_\lambda : \partial\rm  M_{ferm} / 
\partial m_\lambda : \partial M_{\rm s} / \partial m_\lambda  = 5:6:7
\end{equation}

At this point it is worthwhile to return to the question of glueball
masses and their interaction with the S bound states. We make
an observation based on some results of West's  on glueball masses
in QCD \cite{West}. Using QCD inequalties, West has shown that the 
mass of the lightest non-vacuum state coupling to the operator 
$G_{\mu \nu}^2$ is less than the mass of the lightest non-vacuum
state coupling to $G \tilde{G}$. In QCD, this implies that the
lightest glueball is a scalar, not a pseudoscalar. West's results 
can be applied to SYM due to the positivity of the gluino determinant
\cite{Hsu}. In SYM the results are relevant both to the glueballs and
the gluino-gluino bound states, as they have the same quantum numbers
and can mix even at the perturbative level. Now, if (\ref{mass}) 
is correct, then very close to the SUSY point the shift in S bound state
masses is such that the pseudo-scalar S bound state is {\it lighter}
than the scalar S bound state. This is the case for sufficiently small
$m_{\lambda}$ since the unkown K\"{a}hler terms are higher order
in $m_{\lambda}$, and the superpotential splitting dominates. Combining 
this with West's result suggests that in SYM the
glueball states may actually be lighter than the S states near the
SUSY point! (In other words, the S states cannot be dominating West's 
inequalities.) 
If so, lattice measurements of properties of the 
$\lambda \lambda (x)$ correlation functions will actually be dominated
by those light glueballs (due to the mixing), 
and we will not obtain any information on the
VY model. Of course, another possibility is that the glueball--S mixing
is so strong that the VY model is not a very good description in any 
case.

Finally we note, as pointed out in \cite{Shifman},
that the VY model apparently has an extra SUSY vacuum corresponding to
$\langle \phi \rangle = 0$. At this point the expectation value of $S$
is singular and so it is not clear how to interpret this vacuum. Shifman
and Kovner have proposed that the vacuum is real and represents some
conformal, $Z_{2 N_c}$ preserving 
point of the theory. It would be interesting to look for this
vacuum in lattice simulations but unfortunately as can be seen from
Eq.(\ref{newpot}) there is no value of soft breaking mass for which such a
vacuum would be the global minimum. This will make it difficult to
observe in lattice simulations. We return to a possible lattice signature
of this vacuum in the following section.

\section{Tuning to SUSY}

In this section we discuss the problem of tuning a lattice simulation
toward the SUSY limit. As in the case of chiral symmetry in QCD, the
lattice regularization explicitly violates SUSY \cite{CV}. Gauge
invariance only allows a single renormalizable operator that breaks
SUSY, the gaugino mass. One must
tune the bare gaugino
mass to a special value in order to ensure that the
symmetry (in this case SUSY) is restored, and 
even then it is only fully recovered 
in the continuum limit. It is important to identify a sensitive diagnostic
with which to fine-tune the bare gaugino mass. 
(Of course the mass splittings described in the previous section are
a possibility, but they are comparatively slowly varying with $m_\lambda$.)
This is particularly true,
as we will discuss below, because one of 
the most interesting quantities we wish to compute is the gaugino
condensate  $\langle \lambda \lambda \rangle$ which depends sensitively
on small SUSY-breaking effects. We are tempted to 
propose the partition function itself as a diagnostic, 
since it must approach unity in the SUSY limit, 
where the vacuum energy, $\epsilon_0$, is zero,
\begin{equation}
Z ~=~ exp[ - V \epsilon_0] ~ \rightarrow 1~~~.
\end{equation}

The behavior of the vacuum energy is extremely sensitive
to the SUSY limit. In a continuum calculation, one obtains
an expression of the form
\beq
\epsilon_ 0 ~=~ 0 ~+~ m^2 a^{-2} ~+~ m \Lambda^3
~+~ \Lambda^4 {\cal O} ((\Lambda a)^k)~.
\eeq
Here $m$ is the effective gaugino mass, which vanishes when the 
bare gaugino mass $m_\lambda$ is properly chosen. 
When the gaugino mass is non-zero, 
the vacuum energy
receives quadratically divergent corrections, resulting from 
non-cancellation
of gaugino and gauge vacuum loops. 
Since the (bare) gaugino condensate can be related to the
derivative of $Z$ with respect to $m_\lambda$, it too is very sensitive to
SUSY breaking corrections, which will appear multiplied by
two powers of the UV cutoff. If $m$ is not tuned sufficiently close
to zero, the resulting contamination will preclude a measurement
of the non-zero condensate which remains in the SUSY limit.

Unfortunately, due to the nature of the lattice regularization,
the vacuum energy $\epsilon_0$ does not actually approach zero,
even in the SUSY limit. This is because SUSY can only be recovered
in the IR and never in the UV part of the lattice model, 
which nevertheless
contributes to $\epsilon_0$. This is easy to see, because the lattice
dispersion relations of fermions and bosons differ significantly in
higher orders of $(k a)$. SUSY cannot hold
even approximately for modes with momentum $k \sim a^{-1}$.

A better method for identifying the point in parameter space 
corresponding to supersymmetry has been proposed by Montvay \cite{chat}.
As can be seen from Eq.(\ref{deltaV}) the shift in energy of the $N_c$ SYM
vacua depend on the phase of the gaugino mass in the softly broken
theory. The lattice simulations are restricted to real mass terms 
in order to allow Monte Carlo techniques to be employed but
the mass can be tuned through zero to negative values. For even values
of $N_c$ the $n=0$ and $n=N_c/2$
vacua interchange as the true minima when the sign of $m_\lambda$ is flipped.
For odd $N_c$ the $n=0$ vacua is prefered for positive masses, while a
negative real mass places the system on the edge of a 
first order phase transition between two vacua with conjugate phases.

We can use the phase transition between the different
${\rm Z_{N_c}}$ vacua as a rough indicator of SUSY. This transition 
occurs near $m=0$ (i.e. as the mass term switches sign), although
there will be a slight overshoot due to supercooling. 
(For $\vert m \vert$ sufficiently small the critical bubble 
necessary for the transition will be larger than the lattice volume.)
Note that the
sign of the overshoot depends on the direction from which $m=0$ is
approached (hysteresis)\footnote{For the system to exhibit hysteresis,
some ``memory'' of the path in $m$ is required. This would be the
case if, as $m$ is varied, the previous configurations are used 
as the initial ones for the subsequent Monte Carlo update. This allows
the system to be trapped in a metastable phase for
$m_\lambda$ sufficiently close to $m_{\lambda}^*$.}.
We can define the corresponding phase transition points as
$m_\lambda ({\pm})$, and average them to obtain the true SUSY point:
\beq
m_\lambda^* 
~= ~( m_\lambda (+) ~+~ m_\lambda (-) ) / 2,
\eeq
where $$ m( m_\lambda^*  )~ =~ 0.$$
In practice, $m_\lambda ( \pm )$ should be defined as the points where some 
specified order
parameter deviates by some specified amount from its behavior in the
pure phase. A possible order parameter is the Wilson loop, 
$\langle W \rangle$, which is almost independent of $m$ for small $m$, and
hence will only display jumps near the transition points.

It is possible that the existence of the extra vacuum proposed in
\cite{Shifman} and discussed above can be ascertained
by the behavior of the Wilson loop near the phase transition.
Consider the average of a large Wilson loop: $\langle W \rangle$.
We expect that, due to confinement, this exhibits area law behavior
with some string tension $\kappa$:
$$
\langle W \rangle ~ \sim ~ \kappa A ~+~ \cdots~~~,
$$
where the ellipsis indicate subleading effects which scale like
the perimeter of the loop. Note that since the gluinos are in the
adoint representation they, like the gluons, cannot fully screen
sources in the fundamental representation, which must be connected
by gluonic strings.

Now consider what happens to $\langle W \rangle$ at the phase transition,
near $m = 0$. If the system only exhibits the two vacua with non-zero
gluino condensate 
( $\langle \lambda \lambda \rangle ~=~\pm 32 \pi^2 \Lambda^3$ ),
then the important configurations near the transition will consist
of regions of each of these phases, with domain walls in between.
In the bulk of each region the string tension will be essentially 
$\kappa$, assuming that the supercooling is negligible and 
$\vert m \vert$ is very small (this requires a large lattice).
On each Euclidean time slice of the loop, one can imagine the gluonic
string connecting the sources at either end, but passing through
domain walls and regions of each phase in between.
The area law part of $\langle W \rangle$ will remain
the same even at the transition, although the subleading perimeter
effects may exhibit a discontinuity due to the sudden appearance
of the domain walls. Therefore we do NOT expect a discontinuity in the
leading behavior of the Wilson loop.

On the other hand, if a phase exists with 
$\langle \lambda \lambda \rangle ~=~ 0$, as has been suggested in 
\cite{Shifman}, it would presumably have very distinct properties
including a string tension $\kappa' \neq \kappa$. In this case,
at the phase transition the dominant configurations will contain
all {\it three} phases, and the effective string tension will be
altered to some value in between $\kappa$ and $\kappa'$. This
would manifest itself as a very rapid change in the leading
behavior of $\langle W \rangle$. An alternative method of 
searching for the $\langle \lambda \lambda \rangle
= 0$ vacuum is to use the "multicanonical" method, which 
induces transitions
between the different vacua. If the simulation is tuned 
sufficiently close
to the SUSY point, it will then spend a significant portion 
of its time
in the exotic phase.

Also worth investigation is the difference in vacuum 
energy between
the metastable and stable vauca, defined by
\beq
\Delta V (m_\lambda ) ~\equiv~ 
\epsilon_0^{ms} (m_\lambda ) ~-~ \epsilon_0^s (m_\lambda ) ~~. 
\eeq
$\Delta V$ is finite, and vanishes
at the exact SUSY point. The behavior of $\Delta V (m_\lambda )$
is predicted by (\ref{deltaV}), which we emphasize follows directly from
the (Seiberg) effective Lagrangian (\ref{SZ}), independent of the further
V-Y analysis concerning the bound states' masses \cite{VY}.
Furthermore, the slope of $\Delta V$ is directly related to the gaugino
condensate. Alternatively, one could also directly compute the quantity 
\beq
\lim_{m_\lambda \rightarrow 0}~ 
\left[ ~\langle \lambda \lambda \rangle^{ms} ( m_\lambda ) 
~-~ \langle \lambda \lambda \rangle^{s} ( m_\lambda ) ~\right]  ~~~ ,
\eeq 
in which the 
divergences also cancel. 
A sufficiently accurate measurement would also
provide a determination of $\alpha$, as can be seen
from Eq.(\ref{Stau}).

Finally, we mention the possibility of measuring the surface tension
of domain walls separating two different phases. A domain wall 
configuration can be produced by splitting the lattice
into two separate regions, each treated with different 
(say, opposite) values of $m_\lambda$. Recently, Shifman and collaborators
have given exact expressions for the profiles and energy densities of
such configurations, using the $N=1$ SUSY algebra with central extension
\cite{SDW}.

A subtlety in the preceding discusion is that our expressions are given in
terms of the bare quantities in the ``Wilsonian'' regularization scheme
(e.g. $\rm \bar{DR}$ \cite{Pouliot}),
whereas in the lattice simulations it is 
the bare lattice parameters which are varied. Fortunately, one can
relate both quantities to the bare parameters in the usual 
$\rm \bar{MS}$ scheme using perturbation theory.

\section{Summary and Prospects}

To summarize, we believe the following tests of the (softly broken) 
SUSY results will be feasible:

\noindent (1) The first essential step is to confirm the phase structure
of the theory. As discussed in the introduction it is possible that
the SUSY phase ceases to exist for any non-zero gaugino mass, 
although one expects that it will persist until
some critical value of $m$, possibly $m \sim \Lambda$ 
if the gaugino plays an important role in the dynamics of the SUSY theory. 
Assuming that the SUSY phase persists, it should be
described by the softly broken theory and one would hope to 
test the ${\rm Z_{N_c}}$ vacuum structure. This structure
implies the existence of a phase transition near the SUSY point
identified above. There is even the possibilty of using the Wilson loop
$\langle W \rangle$ as a probe of Shifman's putative vacuum at zero
gaugino condensate \cite{Shifman}.

\noindent (2) Measurement of the gaugino condensate and fundamental
scale $\Lambda$. This would provide a direct test
of Seiberg's tower of SQCD superpotentials. 
The shift in the gaugino condensate with increasing $m_\lambda$ 
could also be measured and
would determine the parameter $\alpha$.

\noindent (3) Measurement of mass-splitting ratios $\bar{m}$ (\ref{bm}), 
which
are independent of the relationship between different regularizations.
This is a test of the V-Y Lagrangian itself. At the SUSY point, the
masses should be degenerate and provide a measurement of the
product $\alpha \Lambda$.

\noindent (4) On a more speculative note, some of the recent results
of Shifman and collaborators \cite{Shifman,SDW} are potentially
testable, including the controversial existence of a 
${\rm Z_{2 N_c}}$--preserving vacuum and the exact calculation of
domain wall energy densities.

Finally, we mention a related theory which has
been speculated to have very different behavior from normal QCD 
and which could be easily
simulated using current lattice technology. The model is an SU(2)
gauge theory with a {\it Dirac} spinor of adjoint fermions. This model
was studied in \cite{Banks} with real Susskind fermions in the strong
coupling expansion where it gives rise to a massless composite 
Dirac fermion. This theory can be reached 
at tree level by the inclusion of a large soft
breaking scalar mass in N=2 SQCD, although in that case the
spurion symmetries are not sufficient to determine the IR behavior, so
the results of \cite{Banks} cannot be confirmed without further analysis.
\vspace{.5in}

\noindent {\bf  Acknowledgements}: The authors would like to
thank I. Montvay for detailed discussions. NJE is grateful to the
DESY Laboratory for their hospitality while some of this work was
carried out. SDHH acknowledges Tokyo Metropolitan University
and the Yale-TMU exchange program for its support. MS is grateful
to the Isaac Newton Institute for their hospitality while  a part of this
work was done.
The authors would also like to thank 
R. Sundrum and E. Poppitz for useful input. This work was in part
funded under DOE contracts DE-AC02-ERU3075 and DE-FG02-91ER40676. 

\newpage
\mysection{Appendix: Rescaling Anomaly}

The results given in this paper implicitly assumed the
holomorphic normalizations of fields in the bare Lagrangian.
A rescaling is necessary in order to compare our predictions to 
lattice results which result from canonically normalized bare Lagrangians
(i.e. unit kinetic energy coefficients for the gluino, as opposed to
$1/g^2$). Due to the rescaling anomaly, some care is necessary in this
rescaling \cite{Mur}.

The vector superfields in the holomorphic and canonical normalizations
are related by
\beq
V_h = g_c V_c~~,
\eeq
which, in particular, implies $\lambda_h = g_c \lambda_c$. Naively, using
classical rescaling, we would have the following relation between the
condensates:
\beq
\langle \lambda_h \lambda_h \rangle ~=~ 
 g_c^2 \langle \lambda_c \lambda_c \rangle~~.
\eeq
However, the rescaling anomaly introduces additional effects. In the
holomorphic computation, we essentially computed
\beq
\label{PI}
\int DV_h ~ \lambda_h \lambda_h ~exp \left( - \frac{1}{16 g_h^2} \int 
d^4 x d^2 \theta ~W_h W_h \right) ~~.
\eeq
Changing variables to $V_c$, we have
\beq
\label{RPI}
\int D(g_c V_c) ~ g_c^2 \lambda_c \lambda_c~ 
 exp \left(  - ( \frac{1}{g_h^2}  - \frac{N_c}{4 \pi^2} \ln g_c )    ~   \frac{g_c^2 }{16} \int d^4x d^2 \theta ~W_c W_c \right) ~~,
\eeq
where the shift in coupling by $\frac{N_c}{4 \pi^2} \ln g_c$ is due to
the rescaling anomaly, arising from the functional measure. We see that
if we take 
\beq
\frac{1}{g_h^2} ~=~ \frac{1}{g_c^2}  + \frac{N_c}{8\pi^2} \ln g_c^2~ , 
\eeq
the path integral in (\ref{RPI}) is canonically normalized, with coupling
constant $g_c$.

Our prediction for the canonically normalized lattice
calculation is therefore
\beq
\langle \lambda_c \lambda_c \rangle_{\rm ~lattice, g_c}
~=~  -\frac{1}{g_c^2} 32 \pi^2 \Lambda^3~,
\eeq
where $\Lambda$ is the holomorphic strong couping scale. 
$\Lambda$ is defined by the RGE evolution
of $g_h$, governed by the exact {\it one loop} beta function, 
and is distinct from the strong coupling scale associated
with $g_c$.

%\newpage
%\vskip 0.5in


\baselineskip=1.6pt
\begin{thebibliography}{99}
%
%abbreviated journal names:
%
\def\np#1#2#3{  {Nucl. Phys. #1} (19#3) #2}
\def\pl#1#2#3{  {Phys. Lett. #1} (19#3) #2}
\def\pr#1#2#3{   {Phys. Rev. #1} (19#3) #2}
\def\prep#1#2#3{ {Phys. Rep. #1} (19#3) #2)}
\def\prl#1#2#3{ {Phys. Rev. Lett. #1} (19#3) #2}
%

\bibitem{seiberg1}I.~Affleck, M.~Dine and  N.~Seiberg,
\np{B 241}{493}{84}; \np{B 256} {557}{85}.

\bibitem{seiberg2}  N.~Seiberg, hep-th/9402044, \pr{D 49}{6857}{94}
% EXACT RESULTS ON THE SPACE OF VACUA OF
%FOUR-DIMENSIONAL SUSY GAUGE THEORIES

\bibitem{seiberg3} N.~Seiberg, hep-th/9411149,  \np{B 435}{129}{95}.
%ELECTRIC - MAGNETIC DUALITY IN
%SUPERSYMMETRIC NONABELIAN GAUGE THEORIES.

\bibitem{Montvay} G. Koutsoumbas and  I. Montvay, hep-lat/9612003,
                    \pl{B 398}{130}{97};  I. Montvay,  hep-lat/9607035
                    Nucl. Phys. B 53, Proc. Suppl. (97) 853.

\bibitem{CV}  G. Curci and  G. Veneziano, \np{B 292}{555}{87}.

\bibitem{giradello} L.Giradello and M.T. Grisaru, \np{B 194}{65}{82}.


\bibitem{soft1}  N.~Evans, S.D.H.~Hsu and M.~Schwetz, hep-th/9503186,
\pl{B 355}{475}{95}.
%Exact Results in Softly Broken Supersymmetric Models

\bibitem{soft2} N.~Evans, S.D.H.~Hsu, M.~Schwetz and S.~Selipsky,
hep-th/9508002,  \np{B 456}{205}{95} ; N.~Evans, S.D.H.~Hsu and 
M.~Schwetz, hep-th/9608135, \np{B 484}{124}{97}; hep-th/9703197;
L.~\'{A}lvarez-Gaum\'{e}, J.~Distler, C.~Kounnas and M.~Mari\~{n}o, 
hep-th/9604004, 
Int. J. Mod. Phys. A$11$ (1996) 4745; L.~\'{A}lvarez-Gaum\'{e} and
M.~Mari\~{n}o, hep-th/9606191, Int. J. Mod. Phys. A12 (1997) 975;
L.~\'{A}lvarez-Gaum\'{e}, M.~Mari\~{n}o and F. Zamora, 
hep-th/9703072, hep-th/9707017.
%EXACT RESULTS AND SOFT BREAKING MASSES IN SUPERSYMMETRIC GAUGE  
%THEORY

\bibitem{soft3}  O.~Aharony, J.~Sonnenschein, M.~E.~Peskin,
  S.~Yankielowicz, hep-th/9507013,  \pr{D 52}{6157}{95};
E.~D'Hoker, Y.~Mimura, N.~Sakai, hep-ph/9611458, \pr{D 54}{7724}{96};
K. Konishi,  hep-th/9609021, \pl{B 392}{101}{97}.

\bibitem{VY}
G.~Veneziano and S.~Yankielowicz
\pl{B 113}{231}{82}

\bibitem{softVY} A. Masiero and G. Veneziano, \np{B249}{593}{85}.

\bibitem{Cordes} S. Cordes, \np{B 273}{629}{86}.

\bibitem{Pouliot} D. Finnell and P. Pouliot, hep-th/9503115,
\np{B 453}{225}{95}.

\bibitem{West} G. West, hep-ph/9608258; \prl{77}{2622}{96}, hep-ph/9603316.


\bibitem{Hsu} S.D.H. Hsu, hep-th/9704149.

\bibitem{Shifman} A. Kovner and M. Shifman, TPI-MINN-97-03-T, 
hep-th/9702174.

\bibitem{chat} Private communication with I. Montvay.

\bibitem{SDW} G. Dvali and M. Shifman, \pl {B 396} {64}{97};
A. Kovner, M. Shifman and A. Smilga, TPI-MINN-97/08-T, 
hep-th/9706089.

\bibitem{Banks} T. Banks and V. Kaplunovsky, \np{B 192}{270}{81}.

\bibitem{Mur} N. Arkani-Hamed and H. Murayama, hep-th/9705189
and hep-th/9707133, and references therein.

\end{thebibliography}
\end{document}